\documentclass[]{aa}

\usepackage[dvips]{graphicx}
\usepackage{amsmath}
\usepackage{amssymb}

\begin{document}

\title{The structure of H$_{2}$O shells in Mira atmospheres} 
\subtitle {Correlation with disk brightness distributions and a 
spectrophotometric signature}
\author{A. Tej\inst{1}, A. Lan\c{c}on\inst{1} \& M. Scholz\inst{2}}

\offprints{A. Tej}

\institute{
 Observatoire Astronomique de Strasbourg,
 Universit\'e L.\,Pasteur \& CNRS (UMR 7550), Strasbourg,
France;\\
 \email{surname@astro.u-strasbg.fr}
\and
Institut f. Theoretische Astrophysik der Universit{\"a}t Heidelberg,
Tiergartenstr.15, 69121 Heidelberg, Germany, and School of Physics,
University of Sydney, NSW 2006, Australia;\\
 \email{scholz@ita.uni-heidelberg.de}
}

\date{Received 19 November 2002 / Accepted 6 January 2003}

\authorrunning{A. Tej, A. Lan\c{c}on \& M. Scholz}
\titlerunning{H$_{2}$O shells in Miras}
\abstract{
Dynamic models of M-type Mira variables predict the
occurrence of water ``shells", i.e. of zones of high H$_2$O 
density and high H$_2$O absorption inside the stellar atmosphere. 
The density, position and width of these 
shells is closely correlated with different types of
two-component shapes of the intensity distribution on the 
disk in the $H$, $K$ and $L$ near-continuum 
bandpasses. 
We investigate these correlations and highlight the
role of a spectrophotometric H$_2$O index that warns against serious
complications in diameter measurements
in the case of substantial water contamination of the bandpass of
observation. Simultaneous spectrophotometric and interferometric
measurements may allow observers to estimate real continuum diameters
more precisely.
}

\maketitle

\section{Introduction}
Studies of Mira variables are of considerable astrophysical
interest as these objects are pulsating stars undergoing rapid
mass loss and hence play an important role in the enrichment of the
interstellar medium. Despite substantial progress in recent
years, many fundamental aspects of the physics and structure of Miras still
remain unclear. A summary of the present state of knowledge and of
crucial problems that have to be solved was given, e.g., in a recent review of
Scholz (2002 = S02). Among the most urgent problems is the
determination of the stellar diameter that is needed for 
assigning to the star an effective
temperature $T_{\rm eff}$, for placing the star in the HR
diagram, and for understanding pulsation.

Since the pioneer observations of monochromatic radii of $o$ Cet
and R Leo by Labeyrie and co-workers (Bonneau \& Labeyrie 1973; Blazit et
al. 1977; Labeyrie et al. 1977; Bonneau et al. 1982) and the first studies
of dynamic Mira model atmospheres (Scholz \& Takeda 1987; Bessell et al.
1989), it has been known that the atmospheres of M-type Mira variables are
geometrically very extended configurations. Layers of formation of strong
absorption features may be two or more times as distant from the star's
centre as the continuum-forming layers. For such an extended-atmosphere star,
different diameter definitions may be given (Baschek et al. 1991) and the
observed size of the stellar disk depends on wavelength. The most common
definition of ``the" stellar radius, that describes the global dimensions
of the star and is used in modelling the interior and atmospheric structure,
is the Rosseland radius $R$ which marks the position of the layer
at which the Rosseland optical depth equals unity, $R$ = $r(\tau_{\rm
Ross}$=1).

Clearly, the Rosseland radius is not an observable quantity, but
$R$ is usually close to the continuum optical-depth radius
$R_{\rm cont}$ = $r(\tau_{\rm cont}$=1) which
depends only slightly on wavelength in the near-IR regime and
which in turn may be related via model considerations to an observable
intensity radius (e.g. Bessell et al. 1996 = BSW96; Hofmann et al. 
1998 = HSW98; Jacob \& Scholz 2002 = JS02; S02). Substantial 
deviations of Rosseland radii from continuum radii are found 
in very cool M-type Miras in which the Rosseland
extinction coefficient is noticeably affected by strong molecular
absorption in high layers (e.g. HSW98; JS02; S02).
The intensity radius (at continuum wavelengths 
or in an observer's filter passband)
is defined in terms of the shape of the centre-to-limb variation (CLV).
CLV reconstruction from interferometric data is in
principle possible but not yet feasible with presently available
accuracies and baseline coverage. In practice, an intensity
diameter is derived by fitting observed visibilities with model-predicted
visibilities which are characterized by a radius-type parameter.
Since molecular layers in the outer atmosphere significantly
modify the pure-continuum CLVs, the corresponding visibilities 
and the resulting diameters, 
the choice of an appropriate atmospheric model is critical.

Layers of high molecular density may often be considered 
as a ``molecular shell" which is more or less
detached from continuum-forming layers depending on the
temperature and density stratification of the Mira atmosphere.
Indeed, attempts of interpreting near-IR CLVs in terms of H$_2$O 
``shells" (e.g. Mennesson et al. 2002a, b) or visual light curves 
in terms of TiO ``shells" (Reid \& Goldston 2002) are quite successful. 
Though this approach is useful and illustrative,
one must keep in mind that these H$_2$O or TiO shells are not
circumstellar but are embedded in the stellar atmosphere, and
quantitative interpretation of the effects upon CLVs and spectra may only be
achieved in terms of molecule densities within the framework of
the star's dynamic model atmosphere (e.g. Bessell et al. 1989;
BSW96; HSW98; H\"{o}fner et al. 1998; Woitke et al. 1999).
Model-predicted, shell-type molecular layers of water and various
metal oxides were quoted by Scholz (1992; models of Bessell et
al. 1989) and Woitke et al. (1999). ISO and ground-based
spectroscopic observations also point to warm layers of H$_2$O
(e.g. Hinkle \& Barnes 1979; Tsuji et al. 1997; Yamamura et al.
1999; Matsuura et al. 2002) and other molecules.

In this paper we focus on the characteristics of CLV shapes in
the water-contaminated continuum bandpasses in the near-IR.
We concentrate our discussion on the $K$ bandpass (HSW98; rectangular
filter with $\lambda_{c} = 2.195 \mu m, \Delta\lambda = 0.40 \mu
m$) as this is the filter which is most commonly used 
for interferometric observations of Mira variables.
We also relate the different classes of 
CLVs predicted by the models to the time-dependent structure of 
a pulsating star and to observable spectrophotometric properties. 
A simple classification is described in Sect.\,\ref{Classes.sec}.
In Section\,\ref{Shells.sec}, we demonstrate that shells 
of high water vapour densities are found in the outer layers
of most models as a natural consequence of the density and temperature 
stratification. Their location and extent is shown to relate 
to the shape of the CLVs, establishing water vapour as an important actor  on the
scene of angular diameter measurements. In Sect.\,\ref{Spectro.sec},
we show that spectrophotometric indices sensitive to water bands
can in principle warn the interferometric observer against the 
presence of visibility distortions and provide estimated corrections
to angular diameter values derived in the uniform disk approximation.
However, the reliability of these corrections depends on the
precision and the parameter coverage of predicted
model spectra for individual observed stars.
In a forthcoming paper (Tej et al. 2002 in preparation), 
we will provide an extensive comparison 
with empirical optical to near-IR spectra.

\section{Classification of CLVs}
\label{Classes.sec}
In this paper we investigate the CLVs predicted by the models 
of BSW96 and HSW98. 
For the HSW98 models (P, M, O series), the pulsation is driven by
complete self-excited configurations whereas, for the BSW96 models (Z,
D, E series), a conventional piston approach to the sub-atmospheric
layers is used. The density stratifications results from the shock
front driven outflow and the subsequent infall of matter. Non-grey temperature
stratifications are computed in the approximation of local
thermodynamic and radiative equilibrium. 
Computationally smeared-out density discontinuities occur at shock
front positions. We refer the reader to BSW96 and HSW98 for detailed
discussions on the construction of these dynamical models.

\begin{figure*}
\includegraphics[width=1.0\textwidth]{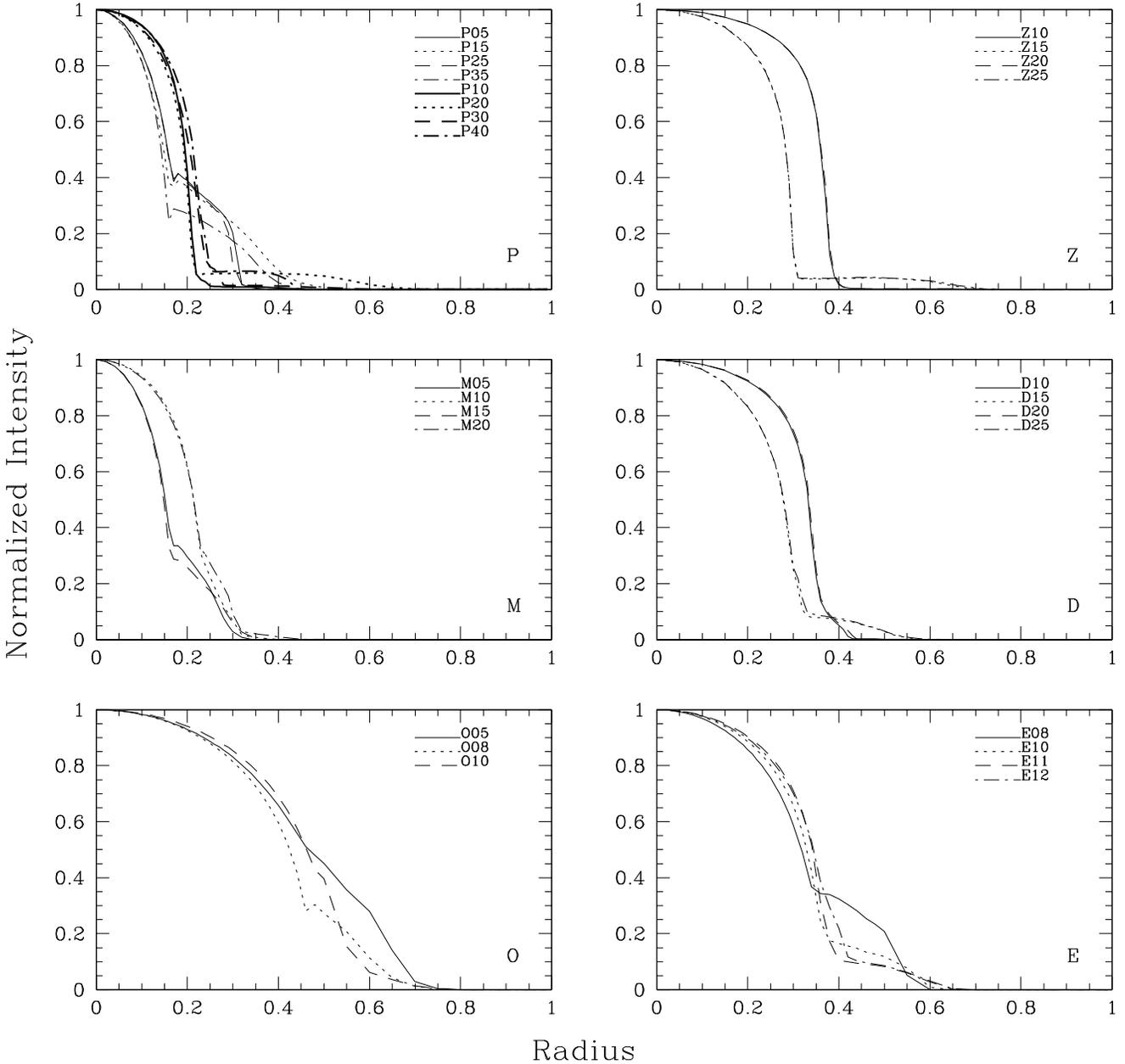}
\caption{The model CLVs. The radius is given in arbitrary units
(consistent within each panel).}
\label{CLVs.fig}
\end{figure*}

The models of Bessell et al. (1989), BSW96 and HSW98 predict a
large variety of CLV shapes. These range from nearly uniform
and moderately darkened disks to complex CLVs with Gaussian-like
or two-component appearance.
The model study of JS02 shows that pure-continuum intensity diameters in
the near-IR may be accurately defined in terms of the position of the steep 
flank of the CLV. The resulting visibilities of 
interferometric observations of the stellar disk may readily be fitted by
a moderately darkened disk, and simple uniform-disk (UD) fits lead to reliable
continuum diameters with errors of the order of only a few percent. This
study also shows, however, that the standard near-IR continuum bandpasses
in the 1.2 to 4.0$\mu$m range ($J$, $H$, $K$, $L$) are noticeably 
contaminated by
atmospheric water absorption in Miras with low effective temperatures and
strongly extended atmospheres. The CLV has a two-component
shape consisting of a moderately darkened inner continuum portion and an
outer tail- or protrusion-type extension. 
The two-component shape of the brightness profile generated in
the Rayleigh-Jeans part of the Planck function by modest
molecular absorption can well be
understood in terms of basic radiation transport (Scholz
2001). The resulting visibility appears
distorted as compared to the pure-continuum visibility. Pronounced 
protrusions may mimic a Gaussian-like shape of the full CLV. 
The detailed discussion of JS02 shows that these effects 
lead to serious consequences for measuring
continuum diameters. Here, we restrict our
investigations and discussions to these two-component CLVs. 
Figure\,\,\ref{CLVs.fig}
displays the model CLVs for the $K$ passband.
The parameter of the non-pulsating ``parent" star and the 
time series of the Mira
models are discussed in HSW98 and quantities relevant to our study are listed in
Table\,\ref{MiraParent.tab} and Table\,\ref{ClassFull.tab}. 
The CLVs can be broadly classified into three
groups based on visual inspection. Note that this is a purely
qualitative description\,;
a more quantitative one will be introduced in Sect.\,\ref{Spectro.sec}.

\begin{table}
\caption{Properties of the Mira ``parent" star series
(taken from BSW96 \& HSW98). The columns: pulsation mode - fundamental (f)
or overtone (o); period $P$; parent star mass $M$; luminosity
$L$; Rosseland radius $R_{\rm p}$; effective temperature $T_{\rm eff} 
\propto\, (L/R_{\rm p}\,^{2})^{1/4}$.}
\label{MiraParent.tab}
\begin{tabular}{l|l|l|l|l|l|l}
\hline
Series & Mode & $P$ & $M$ 
 & $L$ & $R_{\rm p}$ & $T_{\rm eff}$ \\ 
 & & (days) & (M$_{\odot})$ & (L$_{\odot})$ & (R$_{\odot})$ & \\
\hline
Z & f & 334 & 1.0 & 6310 & 236 & 3370 \\
D & f & 330 & 1.0 & 3470 & 236 & 2900\\
E & o & 328 & 1.0 & 6310 & 366 & 2700 \\
P & f & 332 & 1.0 & 3470 & 241 & 2860 \\
M & f & 332 & 1.2 & 3470 & 260 & 2750 \\
O & o & 320 & 2.0 & 5830 & 503 & 2250 \\
\hline
\end{tabular}
\end{table}

\begin{table*}
\caption{Parameters of the time series of the Mira models
(adapted from BSW96 \& HSW98) and classification of model CLVs.
The columns: visual phase $\phi_{vis}$; luminosity $L$; Rosseland radius $R$; 1.04
near-continuum radius $R_{1.04}$ (JS02); effective temperatures
$T_{\rm eff}(R)$ and $T_{\rm eff}(R_{1.04}$); classification and
remarks. 
An abbreviated nomenclature (letter defining the series followed by the
pulsation phase ($\times$ 10)) is adopted for the model names.}
\label{ClassFull.tab}
\begin{tabular}{l|l|l|l|l|l|l|l|l}
\hline
Models & $\phi_{vis}$ & $L/L_{\odot}$ & $R/R_{\rm p}$  & $R_{1.04}/R_{\rm p}$ &
$T_{\rm eff}(R)$ & $T_{\rm eff}(R_{1.04}$) & Type & Remarks \\
\hline
Z10 & 1+0.0 & 7650 & 1.10 & 1.11 & 3350 & 3340 & N & Intensity falls steeply to zero \\
Z15 & 1+0.5 & 3860 & 0.89 & 0.89 & 3150 & 3140 & T & Sharp turn-over \\
Z20 & 2+0.0 & 7750 & 1.11 & 1.12 & 3350 & 3340 & N & Intensity falls steeply to zero \\
Z25 & 2+0.5 & 3830 & 0.89 & 0.89 & 3140 & 3140 & T & Sharp turn-over \\[5mm]

D10 & 1+0.0 & 4490 & 1.04 & 1.04 & 3020 & 3020 & N(?) & Small pedestal \\
D15 & 1+0.5 & 2210 & 0.91 & 0.90 & 2710 & 2720 & T & Smooth turn-over \\
D20 & 2+0.0 & 4560 & 1.04 & 1.05 & 3030 & 3020 & N(?) & Small pedestal as in D10\\
D25 & 2+0.5 & 2170 & 0.91 & 0.90 & 2690 & 2700 & T & Smooth turn-over \\[5mm]

E08 & 0+0.83 & 4790 & 1.16 & 1.07 & 2330 & 2440 & P & Smooth turn-over\\
E10 & 1+0.0  & 6750 & 1.09 & 1.09 & 2620 & 2630 & T & Smooth turn-over\\
E11 & 1+0.1  & 8780 & 1.12 & 1.11 & 2760 & 2770 & T & Smooth turn-over\\
E12 & 1+0.21 & 7650 & 1.17 & 1.15 & 2610 & 2640 & T & Smooth turn-over\\[5mm]

P05 & 0+0.5 & 1650 & 1.20 & 0.90 & 2160 & 2500 & P & Bump \\
P10 & 1+0.0 & 5300 & 1.03 & 1.04 & 3130 & 3120 & T & sharp low intensity \\
P15 & 1+0.5 & 1600 & 1.49 & 0.85 & 1930 & 2560 & P & Bump\\
P20 & 2+0.0 & 4960 & 1.04 & 1.04 & 3060 & 3060 & T & sharp with higher \\
    &       &      &      &      &      &      &   & intensity compared to P10\\
P25 & 2+0.5 & 1680 & 1.17 & 0.91 & 2200 & 2500 & P & Bump\\
P30 & 3+0.0 & 5840 & 1.13 & 1.14 & 3060 & 3050 & T & sharp low intensity\\
P35 & 3+0.5 & 1760 & 1.13 & 0.81 & 2270 & 2680 & P & Bump\\
P40 & 4+0.0 & 4820 & 1.17 & 1.16 & 2870 & 2880 & T & sharp high intensity \\[5mm]

M05 & 0+0.5 & 1470 & 0.93 & 0.84 & 2310 & 2420 & P & smooth turn-over\\
M10 & 1+0.0 & 4910 & 1.19 & 1.18 & 2750 & 2760 & P(?) & smooth turn-over\\
M15 & 1+0.5 & 1720 & 0.88 & 0.83 & 2460 & 2530 & P & smooth turn-over\\
M20 & 2+0.0 & 4550 & 1.23 & 1.20 & 2650 & 2680 & P(?) & smooth turn-over\\[5mm]

O05 & 0+0.5 & 5020 & 1.12 & 1.00 & 2050 & 2130 & P(?) & smooth turn-over\\
O08 & 0+0.8 & 4180 & 0.93 & 0.91 & 2150 & 2170 & P & smooth turn-over\\
O10 & 1+0.0 & 7070 & 1.05 & 1.01 & 2310 & 2360 & P(?) & smooth turn-over\\
\hline
\end{tabular}

{\em Note\,:}\ P(?) indicates protrusions with atypical
characteristics (see text). N(?) indicates 
quasi-normal CLVs, that do however display
a small pedestal.\\
\end{table*}

\begin{itemize}
\item{} Tail Type (T): The intensity of this tail is model dependent.
The turn-over from the flank to the tail feature is usually
located  near or below 0.2 in the normalized intensity scale.
The angle of turn-over also depends on the model.
For instance for P20 the turn-over is sharp but for E10 it is
comparatively smoother. The extent of the tail can also vary as
is seen for P40, where the tail is shorter compared to P20.

\item{} Protrusion Type (P): When the strength of the tail
increases it appears as a protrusion of the inner disk.
The protrusion appears at a high
intensity level. The nature of this feature is also strongly model
dependent. It can either be in the form of a 'bump' as in P35 or
a smoother feature as in M15.

\item{} Normal type (N): These display neither tails nor protrusions
and are not two-component CLVs. They could be modelled
by moderately darkened disks.
\end{itemize}

Table 2 lists the class assigned to each $K$ band CLV, following the above
definitions. The strength of the second CLV component indicates the amount of 
water contamination and is strongly wavelength-dependent, but the basic 
characteristics of the brightness profile are the same in the $H$ and $L$
bandpasses with less and more water, respectively, than the $K$ band (JS02).
In the near-maximum models of the M series, a protrusion-type attachment to 
the inner CLV is seen which is inconspicuous and almost merging with the inner
CLV, in particular in the $H$ band where water absorption is quite modest. 
In the very cool first-overtone O models, the H$_2$O-generated protrusion 
is strong, but it appears only in the near-minimum model O08 as a clearly 
separable component wheareas it blends with the inner CLV in O05 and O10.

It is evident from the plots and the table that a single parameter, 
such as the phase, alone does not explain the presence of tails or protrusions.
For instance, the near-minimum models of the relatively warm Z and D series
have tails whereas the cooler P models have protrusions near-minimum light.
Average temperature and pulsation mode also play a significant role.
If, for the purpose of ordering the classes, we call protrusions
stronger features than tails, there is a clear
correlation between the strength of the feature and the phase
within individual model series. 
The strongest features are found near-minimum. Note that for the overtone
series, which have larger radii and lower $T_{\rm eff}$ than fundamental
pulsators, models show
a global shift of the minimum luminosity to $\sim$ 0.8 (BSW96; HSW98).
In the subsequent sections it
will become evident that the diversity of the CLV shapes seen is due
to a combination of the shock propagation, the shape and extent
of the water shells and the density and temperature
stratification.

\section{Water ``shells" and their implication on CLV shapes}
\label{Shells.sec}

Recent interferometric observations of Miras by Mennesson et al. (2002b)
show a strong increase in the measured UD diameter between the
$K^{\prime}$ and the $L^{\prime}$ bands. 
These authors propose a two-layer scenario with a
classical photosphere and a detached and extended ($\sim$ 3
stellar radii) gas layer to account for this chromatic variation
in size. Similar two-layer models are also invoked by Yamamura et
al. (1999) to model the ISO/SWS spectra of $o$ Cet and Z Cas. 
A two-layer ``slab" model with plane parallel uniform molecular
layers of H$_{2}$O placed one above the other is shown to
accurately fit the observed spectrum.

\begin{figure}
\includegraphics[clip=,width=0.5\textwidth]{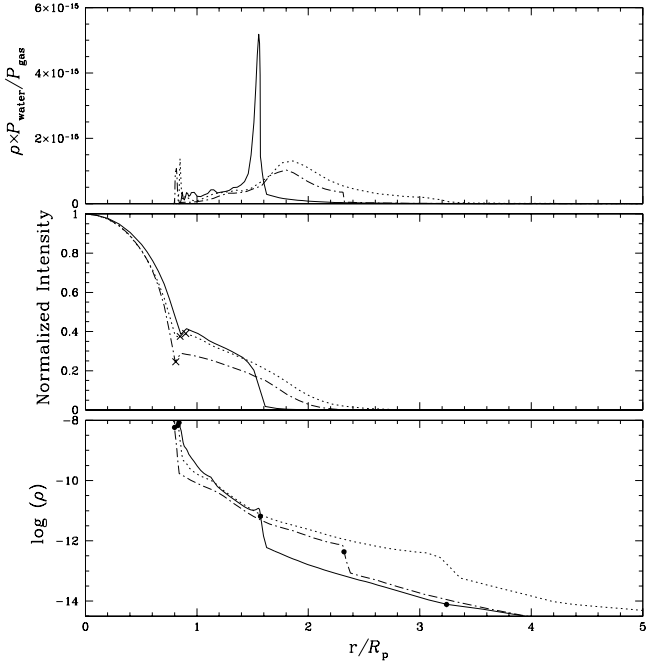}
\caption{The relation between the the water ``shells" (upper
panel), the CLV shapes (middle panel) and the density structure
(lower panel) for the near-minimum P model CLVs. The line
identifications are P05 - solid; P15 - dotted; P35 -
dash-dotted. $R_{\rm p}$ is the radius of the static parent 
star (Table 1). In the middle panel the positions of the
continuum-forming layer ($R_{1.04}$) are shown with crosses.
In the lower panel the location of the shock fronts
(computationally smeared out) are marked with filled circles.}
\label{P_sh_cl_den_prot.fig}
\end{figure}

\begin{figure}
\includegraphics[clip=,width=0.5\textwidth]{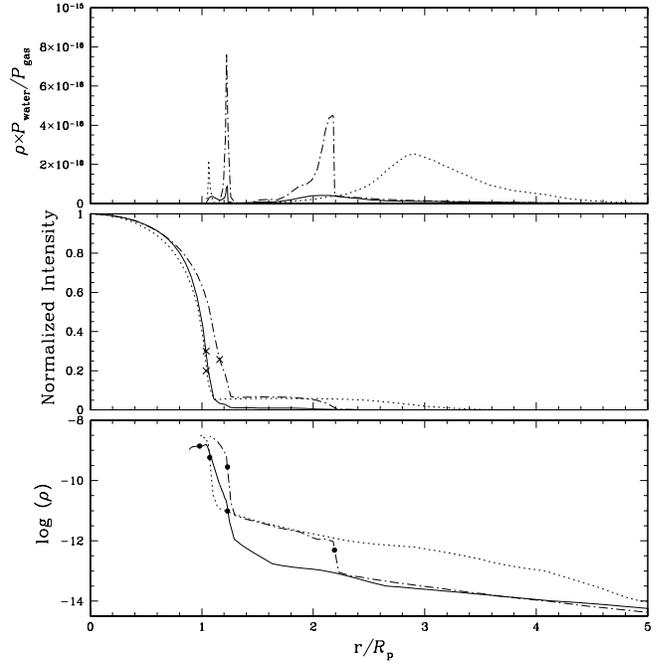}
\caption{Same as in Fig.\,\ref{P_sh_cl_den_prot.fig} for the 
near-maximum models. P10
(solid), P20 (dotted) and P40 (dash-dotted).} 
\label{P_sh_cl_den_tail.fig}
\end{figure}

We investigate the presence of these water shells in time series
of Mira models and their implication on the shape of the CLVs.
To display the shells, we use the quantity
{\it $\rho\times P_{water}/P_{gas}$}, where
$P_{gas}$ is the total gas pressure and $P_{water}$ the partial pressure 
of H$_2$O. It measures the density of water molecules.
In Figures\,\,\ref{P_sh_cl_den_prot.fig} and \ref{P_sh_cl_den_tail.fig},
we plot the P models illustrating the protrusions and tails respectively. 
For Figure\,\,\ref{P_sh_cl_den_prot.fig}, we show the P05,
P15 and the P35 model CLVs (P25 is essentially identical to P05)
and for Figure\,\,\ref{P_sh_cl_den_tail.fig}, we plot the P10, P20
and P40 models (P30 is similar to P40 with a slightly broader
continuum disk).
The upper panels show this shell parameter as a function of the
radius, the middle panels plot the respective CLVs and the lower
panels show the density structure for these models. 
Note that the computational smoothing of the shock fronts
slightly affects the detailed behaviour of the corresponding 
areas in all panels.
The profile of the water shells clearly plays
a role in defining
the shape and strengths of the tail/protrusion component of the CLVs. 

For the protrusion-type P models, which occur
during minima, the effective temperature is low and there is a steep density
gradient after the position of the innermost shock front. Shells are
formed relatively close to the continuum-forming layers. For the P05
model, there is a narrow ``peak-shell" and the density of the
water molecules goes down abruptly at the density discontinuity 
position of the outer
shock front. This manifests itself as a sharp cut-off point on
the protrusion feature beyond which the intensity is small.
Similar behavior is seen for the P35 model where the
density of water molecules sharply decreases at the position of
the outer shock front. The water shell in this case is much
broader and of low density as compared to the P05 model. This
is reflected in the CLV shape where the water component of the CLV
decreases rather slowly. In the P15 model, there is no outer
shock front, hence, the water shell is much more extended. 
The above discussion suggests that the extension of these shells
depends significantly on the nature of propagation of shocks.

The water shells for the tail-type P models are of different nature.
These occur during maxima and for higher effective temperature.
Tails correspond to shells with significantly lower water density
than protrusion. The shells are detached from the continuum-forming
layers and are located in
higher atmospheric regions. P10 has an extended but very-low-density
shell. P20 has a higher density and there is an
abrupt cut-off in the shell coinciding with the position of the
outer shock front. The corresponding CLV has a constant tail
feature up to the cut-off point. The shell is much more extended
for the P40 model and without the occurrence of a outer shock, the
CLV shape remains constant out to larger radii before fading to
zero. 

\begin{figure}
\includegraphics[clip=,width=0.5\textwidth]{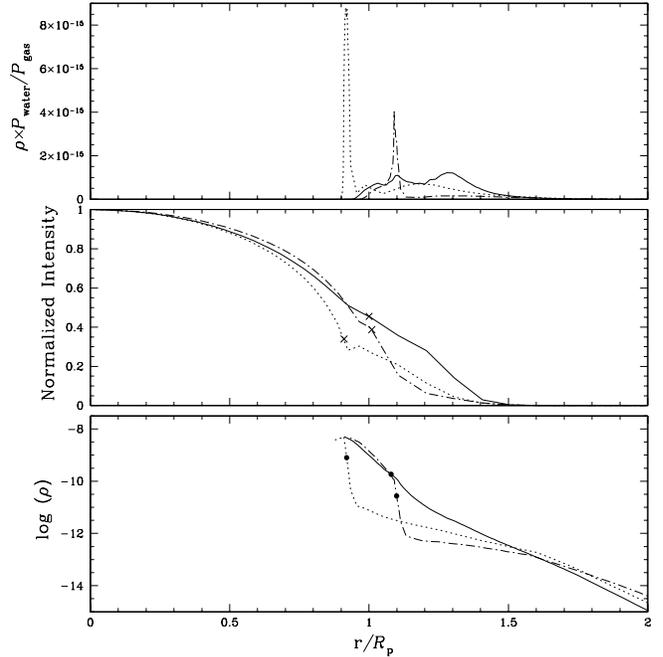}
\caption{Same as in Fig.\,\ref{P_sh_cl_den_prot.fig} for high-mass
overtone
models O05 (solid); O08 (dotted) and O10 (dash-dotted).}
\label{O_sh_cl_den.fig}
\end{figure}

\begin{figure}
\includegraphics[clip=,width=0.5\textwidth]{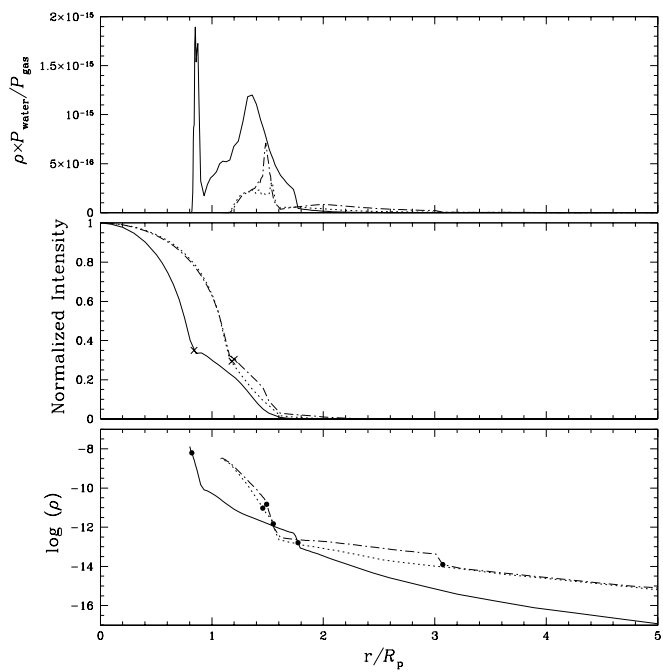}
\caption{Same as in Fig.\,\ref{P_sh_cl_den_prot.fig} for
high-mass fundamental models M05 (solid); M10 (dotted) and M20 (dash-dotted).}
\label{M_sh_cl_den.fig}
\end{figure}

In Figures\,\,\ref{O_sh_cl_den.fig} and \ref{M_sh_cl_den.fig} we
show similar plots for the high-mass overtone O and 
fundamental M models (M15 is similar to M05 and hence not plotted).
As can be seen from all these plots there are
a variety of shapes, strengths and extensions of these water
shells which agree quite well with the variety of shapes seen
in the CLVs. A point worth mentioning here is the presence of a
sharp spike seen in the shells (for e.g. in the P40, O08, M05
models) close to the continuum-forming layers. 
This is due to the enormous inward increase of $\rho$
while $P_{water}/P_{gas}$ is still non-zero. This does not affect
the two-component CLVs but results in some additional darkening of
the normal continuum disk. 

Apart from the very good one-to-one 
correlation seen between the water shells and the shapes of the CLVs, 
in general we see that the shells are detached and more
extended for the tail-type models as compared to the protrusion-type
ones.  This is in gross
agreement with the results of Yamamura et al. (1999) who surmise
that H$_{2}$O layers are more extended at maxima. Mennesson et
al. (2002b) derive a photospheric radius of 10 mas and a H$_{2}$O 
envelope extending from 15 mas to 27 mas for R Leo. This ``gap"
of 1.5 $R_{\star}$ between the continuum layer and the H$_{2}$O
envelope is in good agreement with the shells seen for the tail-type
P models. Occurrence of such ``gaps" is the natural consequence of
the temperature-density requirement of the equation of state
for H$_{2}$O formation.

The density stratification of various molecules 
exhibits shell-like structures at locations similar 
to those of H$_2$O, though with differing 
relative strengths and extents (cf. Woitke et al. 1999). 
For the models used here, JS02 found that molecular contamination of the
$H$, $K$ and $L^*$ continuum bandpasses is essentially due to H$_2$O with
small contributions of OH and CO in the $H$ bandpass. Hence,
molecules other than H$_2$O probably do not appreciably affect
the CLVs and the corresponding visibilities in the near-IR passbands.
However, one should be aware that different filter profiles may
result in different molecular contributions, that the parameter
range and phase-cycle coverage of the models is limited, and
that model assumptions and adopted model opacities may play a
role. For instance, the BSW96 and HSW98 models used here 
assume local thermodynamic equilibrium, whereas Woitke et
al. (1999) have shown significant deviations from equilibrium for
four molecules including H$_{2}$O. Also, variations of
metallicities and of C/O ratios (transition to MS- and S-type
Miras) are not considered in the present models. 


Finally, some interferometric and
spectroscopic observations indicate that inner edges of dust
shells might be as close as 2 or 3 continuum radii from the
star's centre (e.g. Danchi et al. 1994; Danchi \& Bester 1995;
Lobel et al. 2000; Lorenz-Martins \& Pompeia 2000). 
M-type Mira models with thermodynamically
coupled dust (Bedding et al. 2001) show two-component CLVs similar 
to those produced by molecular layers for the 1.04\,$\mu$m bandpass, 
a bandpass that is not significantly contaminated by molecules otherwise.
The distortions are shown to be pronounced around and below 1\,$\mu$m
but the effects are almost negligible for the $H$, $K$ and $L^{*}$
passbands. The situation is likely to depend strongly on the
treatment of dust condensation, and further theoretical work
will be useful. For C-type Mira atmospheres, dust condensation is 
significant in existing models (e.g. H\"ofner et al. 1998),
but these stars lie outside the scope of our paper.

\section{Spectral indications of visibility distortions}
\label{Spectro.sec}

In the previous section we have discussed the occurrence of water
shells, due to absorption in the outer atmospheric layers, 
and their effects on the shape of the CLVs. 
JS02 show that this molecular-band contamination can
distort visibility curves and make the measurement of pure-continuum diameters 
of Miras very difficult.
Visibility distortions are not seen for normal CLVs 
or those with tiny tail-like features (e.g. P10). In such
cases, the diameter estimated by fitting a UD or fully
darkened disk to the measured visibilities would represent the
near-continuum value. But for models which have tails of high
intensity (e.g. P20) the visibilities are highly distorted.
The distortions are even more pronounced and are more
or less Gaussian in shape for models with protrusion-type CLVs (e.g.
P35).
Consequently, the fit diameter (by means of a UD or a
fully darkened disk) is appreciably dependent on the spatial
frequency (i.e. the baseline of observation).
Observationally, this effect is seen
in the shape of high-precision 
visibilities with suitable baseline coverage
(e.g. Perrin et al. 1999; Weigelt et al. 2002) or in the baseline
dependence of UD fit diameters, and in the bandpass-dependence 
of near-continuum UD fit diameters (e.g. Tuthill et al. 2000; 
Mennesson et al. 2002b; Thompson et al. 2002; conference 
reports listed in JS02).

Clearly, the choice of an appropriate atmospheric model and of the 
corresponding CLV is a critical step in the determination
of a stellar diameter. One way of constraining this choice is
to combine interferometric observations with many baselines 
or many wavelengths, and to search a set of atmosphere models for the best
fitting ones. With today's instrumentation and relatively poor
parameter coverage of the models, this remains cumbersome
or impossible. Combining a few interferometric measurements 
with simultaneous spectroscopy should turn out to be easier.
This justifies the search for spectrophotometric indicators
of distorted visibilities.

As it now appears, water molecules are the main culprit
of the two-component nature of the near-IR CLVs.
Hence, we look for spectrophotometric features which 
would quantify the amount of water present in Miras.
The H$_{2}$O index defined by Persson et al. (1977) and Aaronson 
et al. (1978) is a good indicator, that has been frequently used 
for classification purposes (Elias et al. 1982; Terndrup et al. 1991;
Lan\c{c}on \& Rocca-Volmerange 1992). 
It measures the ratio of the fluxes measured through
narrow-band filters centered, respectively, in the wing of the 1.9\,$\mu$m water
band ($\lambda = 2.0\,\mu$m) and in the $K$ band ($\lambda = 2.2\,\mu$m).
The index is expressed in magnitudes and is zero for Vega.
To quantify the strength of the tail/protrusion feature of the
CLVs, we divide the
CLV into two parts - the inner disk, and the tail- or protrusion-
type extension. The ratio of the area of the extension with respect to
the inner disk gives an estimate of the feature strength.

\begin{figure}
\includegraphics[clip=,width=0.5\textwidth]{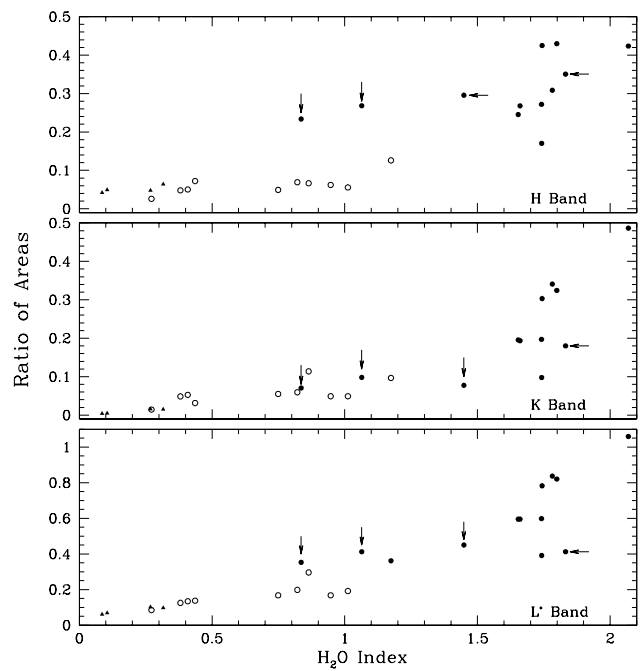}
\caption{Strength of the tail/protrusion
feature as a function of the H$_{2}$O index for the $H$, $K$ and
$L^{*}$ bandpasses. The symbols are: filled triangles - normal
type CLVs; open circles - tail-type CLVs and filled circles -
protrusion-type CLVs. 
The 4 models with atypical P(?)
protrusions in Table 2 (see text) are marked by arrows: M10
(H$_2$O index 0.84), M20 (1.06), O05 (1.83), O10 (1.45).}
\label{HKLh2oclv.fig}
\end{figure}

In Figure\,\,\ref{HKLh2oclv.fig}, we show the feature strength as a
function of the H$_{2}$O index for the $H$ band (HSW98;
rectangular filter, $\lambda_{c} = 1.633 \mu m, \Delta\lambda =
0.30 \mu m$) in the upper panel, the $K$ band in the middle panel 
and the $L^{*}$ band (HSW98; rectangular filter, $\lambda_{c} = 3.799 \mu m, 
\Delta\lambda = 0.60 \mu m$) in the lower panel. 
This figure shows the direct correlation between the strength of
the CLV extensions and the H$_{2}$O index. The strength increases
with the increase in the H$_{2}$O index. For the $H$ and $K$ band CLVs
the ratio of the areas increases slowly from the normal-types to
the tail-types and then there is a steep rise for the protrusion-type
extensions. The three large-ordinate points
marked by arrows in the upper panel of Figure 6 are atypical P(?) protrusions 
in Table 2, which are particularly 
hard to separate in an objective manner from the inner 
CLV for the $H$ band (see Section 2).
Water contamination in the near-IR bandpasses is
stronger in the $L^{*}$ band as compared to the $H$ and $K$
bands (JS02). This manifests in increased
strength of the tail/protrusion feature of the CLVs, as clearly
seen in the lower panel of Figure\,\,\ref{HKLh2oclv.fig}.

The spectrophotometric index of H$_{2}$O gives a satisfactory
estimate of the strength of the tail/protrusion feature of the
CLV. However, the CLVs are not directly observable quantities but
the shapes have a direct implication on the measured visibilities.
JS02 have done a detailed study of the effects of phase, cycle and
baseline on the measured continuum diameters of different Mira
models. They fit UD visibility to the full model-predicted
visibilities by means of a least square routine and also carry out
single point fits of the UD to the model visibility at three
spatial frequencies the choice of which covers the entire range
of various model visibilities. The least square fitting to the
full model takes care of the arbitrary choice of baselines and the
three point fitting gives the baseline dependence of fitted
diameters. The near-IR bands show substantial deviation of the
fitted UD diameters from the near-continuum values. Apart from
uncertainties of model treatment of molecular and dust opacities
at very low temperatures, $R_{1.04}$ (the monochromatic radius
defined at $\tau_{1.04}=1$ in the 1.04 $\mu$m continuum window)
is very close to the continuum value.
We refer the reader to JS02 for a detailed discussion
of the points outlined above. 
As an obvious next step, we explore the correlation
between the H$_{2}$O index and the deviations of fitted UD
diameters from the continuum values. We use the $K$ band results of
Figures 8, 9 \& 10 of JS02 for our study. 

\begin{figure}[ht]
\includegraphics[clip=,width=0.5\textwidth]{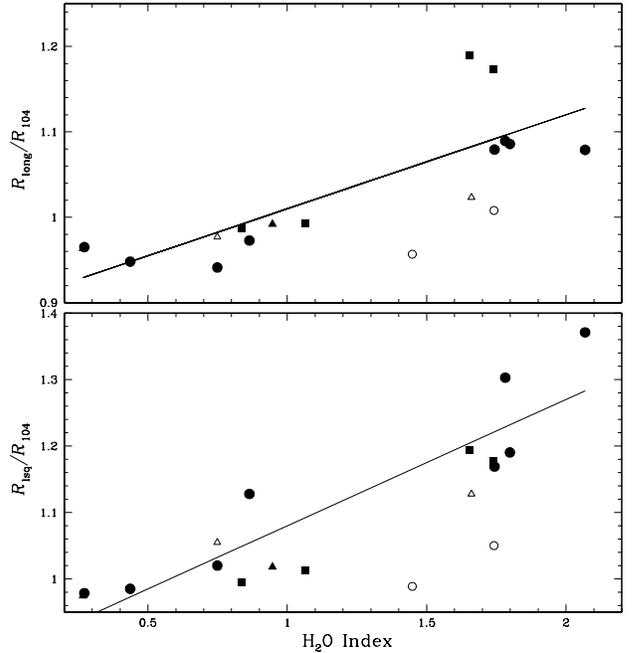}
\caption{In this figure we plot the deviations of the UD fitted
diameters from the the $R_{1.04}$ for the $K$ band. 
The data points are taken from Figures
8,9 \& 10 of JS02. The lower panel shows the results from the
least squares and the upper panel shows the long baseline values.
The symbols are P models - filled circles; D models - filled
triangles; M models - filled squares; O models - open circles; E
models - open triangles}
\label{R_err_104.fig}
\end{figure}
In the lower panel of Figure\,\,\ref{R_err_104.fig}, we
plot the ratio of the least squares fit UD radius ($R_{\rm lsq}$) to
the near-continuum radius ($R_{1.04}$) as a function of the
H$_{2}$O index. Although the dispersion is large, 
the deviations of the fitted UD diameters from the continuum
values distinctly increases with water contamination. 
The relation tightens if the overtone models (O and E series)
are left out. 
A linear least squares fit to the model points excluding
the overtone models yields
\begin{equation}
R_{\rm lsq}/R_{1.04} = 0.19 \,\,{\rm H}_{2}{\rm O} + 0.89
\end{equation}
In the upper panel we plot the correction factor for long baseline observations
($R_{\rm long}$/$R_{1.04}$). This baseline is positioned near the
first null of the visibility function. It is more sensitive to the details of
the CLV structure. 
The least squares fit to this gives
\begin{equation}
R_{\rm long}/R_{1.04} = 0.11 \,\, {\rm H}_{2}{\rm O} + 0.90
\end{equation} 
Similar trends are seen for the
medium and short baselines (not plotted). 
The ratios of interferometric UD radii to the Rosseland radius $R$
(instead of $R_{1.04}$) show much more scattered relations to the
H$_2$O index. This is expected because of the nature of the
Rosseland opacity which is a harmonic-type mean opacity and
is sensitive to extended regions of strong molecular extinction
which tends to close the near-continuum windows, to
increase the Rosseland opacity and, hence, to increase the 
Rosseland radius as compared to the continuum radius.

The above two linear fits emphasize the fact that a simultaneous
spectral observation of water contamination could
yield a correction factor for deriving continuum diameters from
the UD fitted values. A couple of caveats should be kept 
in mind. First, the dispersions in the plots remain
large. The situation should improve when more models become
available, allowing us to explore the origin of the spreads.
For instance, the correlation may depend on
other measurable quantities, such as pulsation properties,
as suggested by the deviant behaviour of the O series.
Second, the exact shape and depth of water bands in spectra
is difficult to compute theoretically, in particular in extended, 
dynamical atmospheres. The reliability of the numerical correction factors 
given above depends on the accuracy the model spectra
and on the range of model parameters explored. As will be
shown in a forthcoming paper, current models are successful
in reproducing the global features of empirical spectra of Miras
and the range of observed water indices, but uncertainties
remain large in the time-dependent depths 
and shapes of individual bands. Once fully satisfactory
models exist, other spectrophotometric measures of the water bands
may turn out to be more sensitive to the radius ratios of 
Fig.\,\ref{R_err_104.fig}.
The series of BSW96 and HSW98 display a tendency for a broader  
1.9\,$\mu$m band in models with protrusions than in models with tails
that remains to be studied in detail.
\section{Conclusion}
In this paper we have classified the two-component near-IR
CLVs predicted by the pulsating Mira models of BSW96 and HSW98 based on
a qualitative visual inspection. The models are shown
to naturally produce water-rich shells depending on
temperature and density stratification of the atmospheric models.
The propagation of shock fronts plays a decisive role in the 
structure and extent of the shells and has direct effect on 
the shapes of the CLVs. The model-predicted strength of the H$_{2}$O
index, quantifying the 1.9 $\mu$m water feature, gives a good
indication of the nature of the CLV in terms of the strength of
the tail/protrusion features with respect to the inner continuum
disk. The increase in water contamination from the $K$ band to the
$L$ band is evident from the increase in the feature strength for
the $L$ band CLVs. 

Using the results of JS02, we show that the
deviations of the fitted UD diameters from the continuum values
are correlated with the H$_{2}$O index, though with a significant 
dispersion. 
The results are based on a moderate number
of models covering a limited range of stellar parameters,
phases and cycles, which prevents us from identifying the
origin of the scatter. 
The exact form of this correlation still has to be determined
from future high-precision visibility observations, and the
usefulness of a simple linear approximation as given in Eqs. (1)
and (2) still has to be checked both empirically and with
more models. Until then,
simultaneous spectroscopic observations of water bands may
at least warn the interferometric observer and provide a rough
estimate of corrections to be applied to the UD fit diameters.


\end{document}